\documentclass[conference]{IEEEtran}
\IEEEoverridecommandlockouts
\usepackage{cite}
\usepackage{amsmath,amssymb,amsfonts}
\usepackage{graphicx}
\usepackage{textcomp}
\usepackage{xcolor}
\usepackage{algorithm}
\usepackage{algpseudocode}
\usepackage{booktabs}  
\usepackage{multirow}     
\usepackage{stfloats}
\usepackage{float} 

\def\BibTeX{{\rm B\kern-.05em{\sc i\kern-.025em b}\kern-.08em
    T\kern-.1667em\lower.7ex\hbox{E}\kern-.125emX}}
\AtBeginEnvironment{equation}{\small}
\AtBeginEnvironment{align}{\footnotesize}
\begin{document}

\title{
	\fontsize{20pt}{20pt}\selectfont
	Post-disaster Max-Min Rate Optimization for Multi-UAV\\ RSMA Network in Obstacle Environments
}

\author{
	Qingyang Wang$^{\dagger}$, Zhuohui Yao$^{\dagger}$, Wenchi Cheng$^{\dagger}$ and Xiao Zheng$^{\dagger}$\\
	\small $^{\dagger}$State Key Laboratory of Integrated Services Networks, Xidian University, Xi’an, China\\
	\small E-mails: wangqingyang@stu.xidian.edu.cn, yaozhuohui@xidian.edu.cn, wccheng@xidian.edu.cn, zheng\_xiao@stu.xidian.edu.cn
	\thanks{This work was supported in part by the Postdoctoral Fellowship Programs of CPSF (Nos. 2023TQ0256 and GZC20232056) and the Fundamental Research Funds for the Central Universities (ZYTS25272).}
}

\maketitle
\begin{abstract}
This paper proposes a rate-splitting multiple access (RSMA) transmission scheme to maximize the minimum achievable rate among ground users for emergency communications in post-disaster scenarios with obstacles, with which the optimal positioning of multiple unmanned aerial vehicle (UAV)-enabled base stations can be achieved timely.
To address the resulting non-convex and intractable optimization problem, we design an alternating optimization approach. Specifically, we relax obstacle-related constraints using penalty terms. In each iteration, block coordinate descent (BCD) and successive convex approximation (SCA) are applied alternately to obtain locally optimal solutions, and penalty multipliers are updated to ensure convergence of the relaxed problem to the original one. Simulation results demonstrate that the proposed scheme significantly outperforms benchmark methods in terms of the minimum achievable rate, verifying its effectiveness and superiority.
\end{abstract}


\section{Introduction}
Unmanned aerial vehicles (UAVs) have been widely adopted in wireless communications due to their unique superiorities on line-of-sight (LoS) link construction and robust capabilities\cite{emergencyUAV}. In specific, UAV can rapidly enhance the deteriorated communication links by providing virtual LoS path and flexibly configure nodes in the system to achieve network robustness, which can well support specific demands for various environments, especially in emergency communications \cite{multi-uav networks}.
In harsh environments after disasters, UAVs can be deployed rapidly to provide temporary communication recovery solutions when the traditional communication infrastructure fails or becomes significantly compromised\cite{emergency Yao}.
Hence, UAV-assisted network is regarded as one of the promising techniques and effective solutions for emergency communications.

To exploit their unique advantages, several important research topics have emerged. For example, research on resource allocation to improve emergency wireless communication quality\cite{Resource Allocation}, or UAV systems focused on deployment aiming at providing fairness service in resource-constrained environments\cite{signal uav and block}. However, these works fails to drastically harness the capability of coordination for UAV network and lacks of accurately capturing the and the inherent features of emergency environments. 
In specific, due to limitations in flight altitude and transmit power, single UAV fails to provide satisfactory service coverage for all users in such challenging environments. Besides, collapsed buildings cause obstacle environments that drastically attenuate communication links.
Therefore, deploying multiple UAVs with well-designed strategies is important for establishing a reliable, wide-coverage and highly adaptable network that not only enhances service quality but also enables desirable LoS communication for users by effectively avoiding weak none-line-of-sight (NLoS) condition. However, meeting the growing demands for spectral efficiency enhancement and inter-UAV interferences management is challenging. Thus, a new communication method with high spectral efficiency, strong interference resistance, and guaranteed minimum rate in emergencies is needed for multi-UAV assisted systems.


In state-of-the-art research, rate-splitting multiple access (RSMA), as a promising non-orthogonal multiple access (NOMA) technique, has empowered UAV-based base stations to address the aforementioned challenges, which show superiority and effectiveness over both space-division multiple access (SDMA) and conventional NOMA schemes\cite{rsma noma sdma}. 
Specifically, a novel interference management scheme can boost the spectrum efficiency of downlink transmission\cite{rsma spectrum}. Owing to the presence of a common stream, each user is ensured a baseline data rate, while private streams are further used to enhance individual performance. This hierarchical transmission structure naturally benefits the maximization of the minimum user rate.

Additionally, the significant gains in UAV emergency communications rely on establishing LoS links, but due to the complex blockage scenario caused by post-disaster environment, it is challenging to consistently maintain communication under LoS conditions.
Existing methods, such as point cloud sensing\cite{point clould} and radio map prediction\cite{radio map}, can assist the LoS link construction, but they are resource-intensive and lack of adaptability. Capturing the unique features of emergency environments and then efficiently design the UAVs' position are the key points to maintain a long-lasting LoS condition to improve the transmission link.

\begin{figure}[t]
    \centering
    \includegraphics[width=0.48\textwidth]{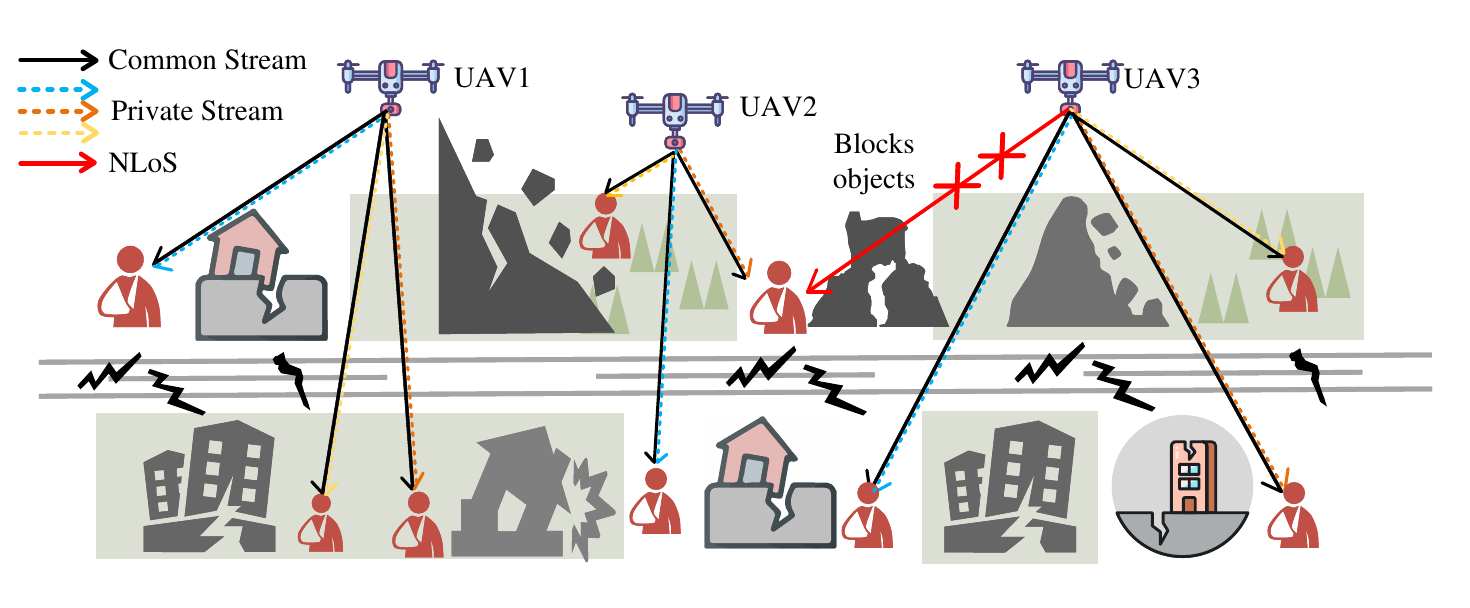}
    \caption{System model.}
    \label{fig:drawing1}
\end{figure}
Motivated by the above facts, in this paper we propose a
 LoS determination approach that utilizes environmental information to assess the spatial relationship among UAVs, users, and obstacles, providing a simple and effective solution for LoS guarantee.
In specific, we propose a multi-UAV positioning strategy integrated with RSMA to enhance communication fairness in obstacle environments. We reconstruct the LoS condition by using the LoS method and jointly optimize UAV positioning, power allocation, and user association under practical constraints. To handle the binary nature of LoS conditions and the problem's non-convexity, we adopt a penalty-based reformulation and solve it using a block coordinate descent (BCD) framework integrated with successive convex approximation (SCA). Simulation results show that the proposed approach significantly improves the minimum achievable rate and outperforms conventional schemes with respect to user coverage and fairness performance.
\begingroup
\section{\small SYSTEM MODEL AND PROBLEM FORMULATION}
\endgroup
As illustrated in Fig. 1, we consider a system model where multiple UAVs act as base stations, denoted by the set $\mathcal{M}$.
The set  $\mathcal{K}$ denotes the ground users served in the network. Meanwhile, obstacle buildings distributed throughout the environment are represented by the set $\mathcal{Q}$.
 all UAVs are assumed to operate on the same frequency band, leading to multi-UAV interference among ground users.
 Besides, the locations of users are randomly and sparsely distributed across the environment with obstacles causing obstructions, making multiple UAVs necessary to effectively handle such deficiency.
  Additionally, UAVs are deployed at higher altitudes than that of the buildings to avoid potential collisions.

Without loss of generality, we employ a 3-D Cartesian coordinate system. For user $k$, its coordinates is denoted by $\mathbf{u}_k \in \mathbb{R}^3$ with $k \in \mathcal{K}$. The coordinates of $m$-th UAV  are given by $\mathbf{x}_m \in \mathbb{R}^3$ with $m \in \mathcal{M}$. We denote by $\mathbf{X}=\{\mathbf{x}_m|m \in \mathcal{M}\}$ the set comprised of all UAVs' coordinates. 
\subsection*{A. Channel Model} 
The channel coefficient between user $k$ and UAV $m$ with antennas number $j$ is modeled as a function of UAV position $\mathbf{x}_m$,
\begin{equation}
	h_{k,m}^j = \frac{\sqrt{\beta_k(\mathbf{x}_m)}}{\| \mathbf{x}_m - \mathbf{u}_k \|^{\frac{\alpha_k(\mathbf{x}_m)}{2}}},
\end{equation}
where $||\mathbf{x}_m - \mathbf{u}_k||$ is the distance between UAV and user, $\alpha \in \{\alpha^{\mathrm{LoS}}, \alpha^{\mathrm{NLoS}}\}$ is the path loss exponent, and $\beta \in \{\beta^{\mathrm{LoS}}, \beta^{\mathrm{NLoS}}\}$ denotes the channel power gain. In order to distinguish between LoS and NLOS channels, the key point is to detect whether there is a blocking effect between the user and the UAV.

From the user's perspective, as shown in Fig.~2, for each building $q$, the one or two side surfaces closest to the user are identified as the visible surfaces. These surfaces have non-base, non-repeated outer edges, referred to as visible outer edges and highlighted in yellow. Each visible outer edge is denoted by $\overrightarrow{L_{k,q,i'}L_{k,q,i''}}$, where $i'$ and $i''$ represent the two vertices of the $i$-th visible outer edge. Together with the user, these vertices define the blocking plane associated with building $q$. The region enclosed by this blocking plane constitutes the user's NLoS space. Drawing inspiration from article \cite{signal uav and block}, we use the directed distance\footnote{\textit{Directed distance}: is the signed distance based on the outward normal vector of the blocking plane reflecting relative position.} from the UAV to the blocking plane to determine whether the UAV lies within the user's NLoS region. 
\begin{figure}[t]
	\centering
	\includegraphics[width=0.4\textwidth]{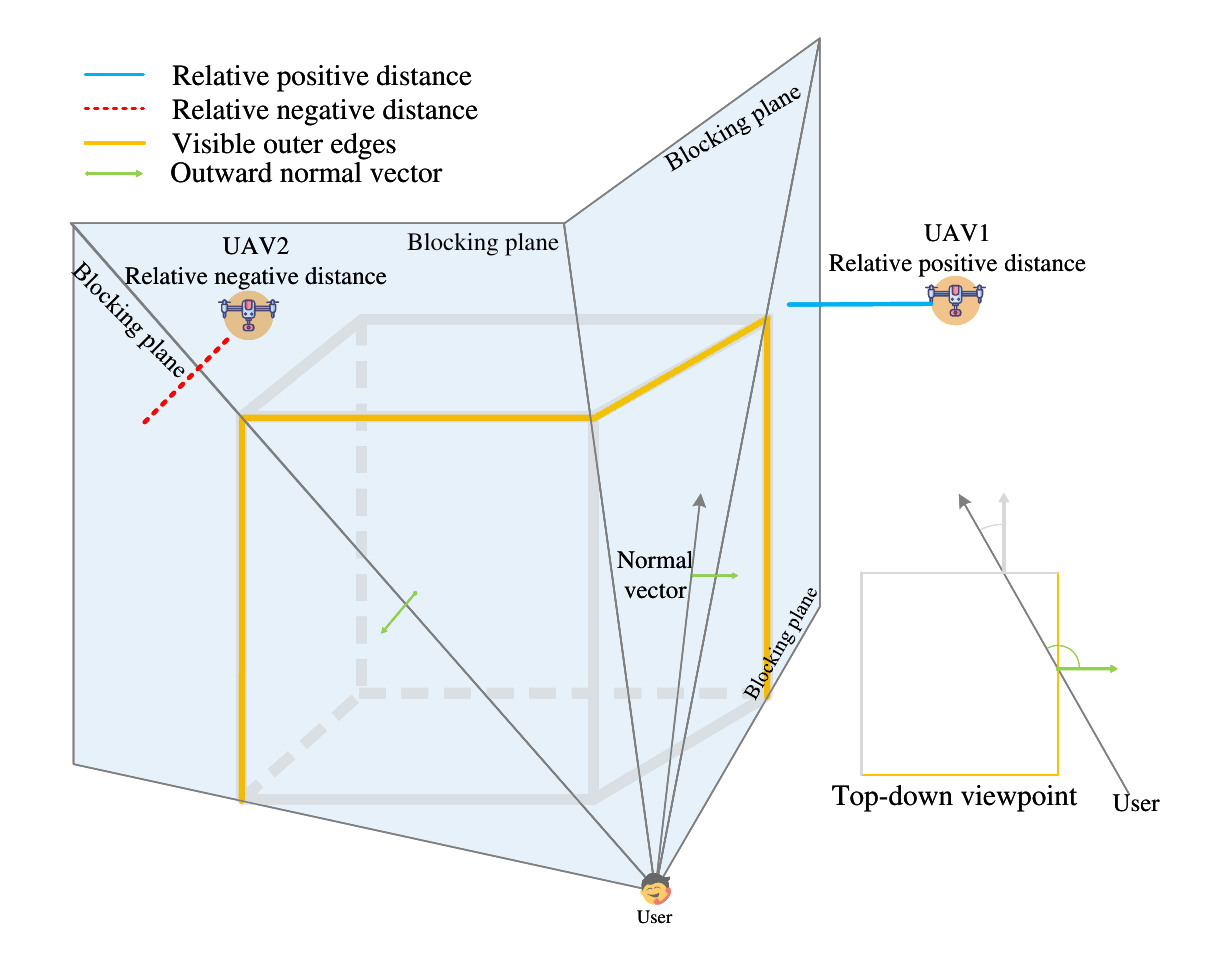}
	\caption{Illustration of the blockage caused by building.}
	\label{fig:drawing2}
\end{figure}
Specifically, the visible surface is defined when the outward normal vector forms an obtuse angle with the vector from the user to building surface. The blocking plane, defined by the user and the visible outer edges, has an outward normal vector denoted by $\mathbf{n}_{k,q,i}=\frac{\overrightarrow{SL_{k,q,i'}}\times \overrightarrow{SL_{k,q,i''}}}{\|\overrightarrow{SL_{k,q,i'}}\times \overrightarrow{SL_{k,q,i''}}\|}$, where $S$ is user's position. The directed distance of origin point and blocking plane $i$ is represented by $d_{k,q,i}^{\text{origin}}=\mathbf{n}_{k,q,i}\cdot\overrightarrow{OS}$. For a UAV located at position $\mathbf{x}_m$, the directed distance to this blocking plane is calculated as:
\begin{equation}
	d_{k,q,i}^{\text{directed}}=\mathbf{n}_{k,q,i} \cdot \mathbf{x}_m - d_{k,q,i}^{\text{origin}},
\end{equation}
if \( d_{k,q,i}^{\text{directed}} \) is negative for all blocking planes associated with building \( q \), the UAV is located in the user's NLoS region, where communication is severely degraded due to obstacles.

In the broader environment, for all buildings, if blocking planes of each building yield at least one positive directed distance, the UAV is considered to be in a LoS condition.
Oppositely, if there exists a building for which all directed distances are negative, the UAV and the user are in a NLoS condition. The expression for the LoS region, denoted by \( \mathcal{D} \), is
\begin{equation}
	\mathcal{D}_k = \left\{ \mathbf{x} \in \mathbb{R}^3 \;\middle|\; \forall q \in \mathcal{Q},\; \left( \exists i \in \mathcal{I}_{k,q} \text{ s.t. } d_{k,q,i}^{\text{directed}} > 0 \right) \right\}.
\end{equation}

UAV \( m \) transmits messages to a subset of users \( \mathcal{K}_m \subseteq \mathcal{K} \). For the $k$-th user with \( k \in \mathcal{K}_m \), its corresponding message \( W_k \) is divided into a common part \( W_{k,m}^c \) and a private part \( W_{k,m}^p \). The common parts \( \{ W_{k,m}^c \mid k \in \mathcal{K}_m \} \) are encoded into a single common stream \( s_{c,m} \), which is decodable by all users in \( \mathcal{K}_m \). Each private part \( W_{k,m}^p \) is encoded into a private stream \( s_{k,m} \). The transmitted stream vector is \( \mathbf{s}_m = [s_{c,m}, s_{1,m}, \ldots, s_{|\mathcal{K}_m|,m}]^T \), which is precoded using the matrix \( \mathbf{P}_m = [\mathbf{p}_{c,m}, \mathbf{p}_{1,m}, \ldots, \mathbf{p}_{|\mathcal{K}_m|,m}] \),
 where \( \mathbf{p}_{j,m} \) is the precoder for stream \( s_{j,m} \) with \( j \in \{c, 1, \ldots, |\mathcal{K}_m|\} \).
\begin{equation}
	\mathbf{x}_m = \mathbf{p}_{c,m} s_{c,m} + \sum_{k \in \mathcal{K}_m} \mathbf{p}_{k,m} s_{k,m},
\end{equation}
where \( E[|s_{c,m}|^2] = 1 \), \( E[|s_{k,m}|^2] = 1 \) $E(\cdot)$ denotes the expectation. The received signal at user \( k \in \mathcal{K} \) from all UAVs, denoted by $y_k$, is:
\begin{equation}
	y_k = \sum_{m \in \mathcal{M}} \mathbf{h}_{k,m}^H \mathbf{x}_m + n_k,
\end{equation}
where \( n_k \sim \mathcal{CN}(0, \sigma^2) \) is additive white Gaussian noise with variance \( \sigma^2 \).

For simplicity, we consider a simplified scenario in the absence of precoder at UAV antennas. The effective channel gain is defined as:
\begin{equation}
g_k(\mathbf{x}_m) = \mathbf{h}_{k,m}^\mathbf{H}\mathbf{h}_{k,m} = \frac{\beta_k(\mathbf{x}_m)}{\|\mathbf{x}_m - \mathbf{u}_k\|^{\alpha_k(\mathbf{x}_m)}}, \\
\end{equation}
the transmit power allocated from the $m$-th UAV to the $k$-th user is defined as:
\begin{equation}
p_{k,m} = \mathbf{p}_{k,m}^\mathbf{H} \mathbf{p}_{k,m}.
\end{equation}

When the $k$-th user with  \( k \in \mathcal{K}_m \) decodes the common stream transmitted by its serving UAV, the part of the common stream corresponding it is the intended signal, while the remain components is regarded as interferences. SINR for the common stream at user \( k \) is given by:
\begin{equation}
	\text{SINR}_{k,m}^c(\mathbf{X}, \mathbf{P}) = \frac{p_{c,m} g_k(\mathbf{x}_m)}{\sum_{j \in \mathcal{K}_m} p_{j,m} g_k(\mathbf{x}_m) + I_{k,m} + \sigma^2},
\end{equation}
where
\begin{equation}
	\small
	I_{k,m} = \sum_{m' \in \mathcal{M} \setminus \{m\}} \left( p_{c,m'} g_k(\mathbf{x}_{m'}) + \sum_{j \in \mathcal{K}_{m'}} p_{j,m'} g_k(\mathbf{x}_{m'}) \right),
\end{equation}
represents interferences to the $k$-th user from other UAVs.

Let \( \mathbf{C} = \{ c_{k,m} \mid k \in \mathcal{K}, m \in \mathcal{M} \} \) represents the user-UAV association matrix, where \( c_{k,m} = 1 \) indicates that $m$-th UAV serves $k$-th user, and \( c_{k,m} = 0 \) otherwise. The achievable rate of the common rate for user \( k \) is:
\begin{equation}
R_{k,m}^c(\mathbf{X}, \mathbf{P}, \mathbf{C}) = c_{k,m} \log_2 \left( 1 + \text{SINR}_{k,m}^c(\mathbf{X}, \mathbf{P}) \right).
\end{equation}

To ensure all users in \( \mathcal{K}_m \) can decode the common stream, the common rate is set as \( R_{m}^c = \min_{k \in \mathcal{K}_m} R_{k,m}^c(\mathbf{X}, \mathbf{P}, \mathbf{C}) \). In order to simplify the algorithm, we assume that this rate is equally allocated among all users served by $m$-th UAV, yielding the per-user common rate:
\begin{equation}
R_{k,m}^{c,\text{avg}}(\mathbf{X}, \mathbf{P}, \mathbf{C}) = \frac{R_{m}^c}{|\mathcal{K}_m|}.
\end{equation}

After successfully decoding and subtracting the common stream, each user \( k \in \mathcal{K}_m \) decodes its private stream  in order of decreasing channel gain $ g_k(\mathbf{x}_m) $. Once a user’s private stream is decoded, its corresponding part is subtracted from the composite stream, leaving interferences only from the latter users with smaller SINR. Thus, the SINR for the private stream is:
\begin{equation}
\small	
\text{SINR}_{k,m}^p(\mathbf{X}, \mathbf{P}) = \frac{p_{k,m} g_k(\mathbf{x}_m)}{\sum_{j \in \mathcal{K}_m, j > k} p_{j,m} g_k(\mathbf{x}_m) + I_{k,m} + \sigma^2}.
\end{equation}

The achievable private rate for $k$-th user is:
\begin{equation}
	R_{k,m}^p(\mathbf{X}, \mathbf{P}, \mathbf{C}) = c_{k,m} \log_2 \left( 1 + \text{SINR}_{k,m}^p(\mathbf{X}, \mathbf{P}) \right).
\end{equation}

The total rate for $k$-th user from $m$-th UAV is:
\begin{equation}
R_{k,m}(\mathbf{X}, \mathbf{P}, \mathbf{C}) = R_{k,m}^{c,\text{avg}}(\mathbf{X}, \mathbf{P}, \mathbf{C}) + R_{k,m}^p(\mathbf{X}, \mathbf{P}, \mathbf{C}).
\end{equation}

The overall rate for $k$-th user is:
\begin{equation}
R_k(\mathbf{X}, \mathbf{P}, \mathbf{C}) = \sum_{m \in \mathcal{M}} R_{k,m}(\mathbf{X}, \mathbf{P}, \mathbf{C}).
\end{equation}
\subsection*{B. Problem Formulation} 
We focus on maximizing the minimum rate to ensure the fairness among all users. We formulate the following problem by optimizing the UAV positioning $\mathbf{X}$, power $\mathbf{P}$, and user association $\mathbf{C}$. The optimization problem is formulated as follows:
\begin{subequations} \label{eq:optimization}
	\begin{align}
		&\max_{\mathbf{X}, \mathbf{P}, \mathbf{C}} \min_{k \in \mathcal{K}} R_k(\mathbf{X}, \mathbf{P}, \mathbf{C}) \label{eq:optimization_a}\\ 
		\text{s.t.} \quad 
		& c_{k,m} \in \{0, 1\}, \quad \forall k \in \mathcal{K}, m \in \mathcal{M}, \label{eq:optimization_b}\\
		& \sum_{k \in \mathcal{K}} c_{k,m} \leq  |\mathcal{K}_m|, \quad \forall m \in \mathcal{M}, \label{eq:optimization_c}\\
		& \sum_{m \in \mathcal{M}} c_{k,m} = 1, \quad \forall k \in \mathcal{K}, \label{eq:optimization_d}\\
		& p_{c,m} + \sum_{k \in \mathcal{K}_m} p_{k,m} \leq P_{\max}, \quad \forall m \in \mathcal{M}, \label{eq:optimization_e}\\
		& p_{c,m} \geq 0, \ p_{k,m} \geq 0, \quad \forall m \in \mathcal{M}, k \in \mathcal{K}, \label{eq:optimization_f}\\
		& \mathbf{x}_m \in \mathcal{D}, \quad \forall m \in \mathcal{M}. \label{eq:optimization_g}
	\end{align}
\end{subequations}
where~\eqref{eq:optimization_b} indicates association is binary.
\eqref{eq:optimization_c} ensures that each UAV can serve $|\mathcal{K}_m|$ users at most. \eqref{eq:optimization_d} guarantees that each user is connected to one UAV. Constraints \eqref{eq:optimization_d} and \eqref{eq:optimization_d} indicate that the common and private transmit power of each UVA is nonnegative
and sum of that not exceed a value $P_{\max}$. Constraint \eqref{eq:optimization_g} indicates the UAVs positioning region.

\section{Optimization Problem}
\subsection*{A. Problem Transformation} 
Note constraint~\eqref{eq:optimization_b} involves binary variables, it is generally hard to tackle directly. Therefore, we convert it into an equivalent form given by:
\begin{subequations} \label{eq:penalty}
	\begin{align}
		& 0 \leq c_{k,m} \leq 1, \quad \forall k \in \mathcal{K}, \forall m \in \mathcal{M}, \label{eq:penalty_a} \\
		& c_{k,m,n} (1 - c_{k,m,n}) \leq 0, \quad \forall k \in \mathcal{K}, \forall m \in \mathcal{M}. \label{eq:penalty_b}
	\end{align}
\end{subequations}

The equivalent form ensures that only binary solutions with \( c_{k,m} \in \{0, 1\} \) are feasible. To simplify the optimization, we relax \( c_{k,m} \) to a continuous variable within \([0, 1]\). Then, constraint ~\eqref{eq:penalty_b} is penalized in the objective using multipliers \( \Lambda = \{ \lambda_{k,m} \mid k \in \mathcal{K}, m \in \mathcal{M} \} \), resulting in a penalized formulation as follows:
\begin{align}
	\underset{\mathbf{X}, \mathbf{P}, \mathbf{C}}{\text{max}} \quad &\underset{k \in \mathcal{K}}{\text{min}} R_k(\mathbf{X}, \mathbf{P}, \mathbf{C}) + \rho(\mathbf{\Lambda}, \mathbf{C})\\
	\text{s.t.} \quad &\lambda_{k,m} \geq 0, \quad k \in \mathcal{K}, m \in \mathcal{M}, \\
	&\eqref{eq:optimization_c},\eqref{eq:optimization_d}, \eqref{eq:optimization_e}, \eqref{eq:optimization_f}, \eqref{eq:optimization_g}, \eqref{eq:penalty_a}\notag.
\end{align}

where $\rho(\mathbf{\Lambda}, \mathbf{C}) \overset{\Delta}{=} -\sum_{m \in \mathcal{M}}\sum_{k \in \mathcal{K}} \lambda_{k,m} c_{k,m} (1 - c_{k,m})$ is the penalty term that penalizes the objective function with $ c_{k,m} \neq 0,1$.

The total power consumed by $m$-th UAV and other interferences for signal transmission can be expressed by:
\begin{equation}
	\footnotesize
	\begin{cases}
		P_{k}^{\text{total}}(\mathbf{X},\mathbf{P}) = \sum\limits_{m \in \mathcal{M}} \left( \frac{p_{c,m}}{\sigma^2}  + \sum\limits_{j \in \mathcal{K}_{m}} \frac{p_{j,m}}{\sigma^2}\right)g_k(\mathbf{x}_{m}), \\
		I_{k,m}^{\text{other}}(\mathbf{X},\mathbf{P}) = \sum\limits_{m' \in \mathcal{M} \setminus \{m\}} \left( \frac{p_{c,m'}}{\sigma^2}  + \sum\limits_{j \in \mathcal{K}_{m'}} \frac{p_{j,m'}}{\sigma^2} \right)g_k(\mathbf{x}_{m'}),
	\end{cases}
\end{equation}
respectively. For notation convenience, we define:
\begin{equation}
	\footnotesize
	\left\{
	\begin{aligned}
		\hat{R}_{k,m}(\mathbf{X}, \mathbf{P}, \mathbf{C}) &= c_{k,m} \log_2 \left( 1 + P_{k,\text{total}} \right), \\
		\bar{R}_{k,m}^c(\mathbf{X}, \mathbf{P}, \mathbf{C}) &= c_{k^*,m} \log_2 \left( 1 + \sum_{j \in \mathcal{K}_m} \frac{p_{j,m}}{\sigma^2} g_{k^*}(\mathbf{x}_m) + I_{k^*,m}^{\text{other}} \right), \\
		\text{where } k^* &= \arg\min_{k \in \mathcal{K}_m} g_k(\mathbf{x}_m), \\
		\bar{R}_{k,m}^p(\mathbf{X}, \mathbf{P}, \mathbf{C}) &= c_{k,m} \log_2 \left( 1 + \sum_{\begin{subarray}{c} j \in \mathcal{K}_m \\ j > k \end{subarray}} \frac{p_{j,m}}{\sigma^2} g_k(\mathbf{x}_m) + I_{k,m}^{\text{other}} \right),
	\end{aligned}
	\right.
\end{equation}
then the achievable rate can be expressed as follows:
\begin{equation}
	\small 
	\begin{cases}
		R_{k,m}^c(\mathbf{X}, \mathbf{P}, \mathbf{C}) = \hat{R}_{k^{*},m}(\mathbf{X}, \mathbf{P}, \mathbf{C}) - \bar{R}_{k,m}^c(\mathbf{X}, \mathbf{P}, \mathbf{C}), \\
		R_{k,m}^p(\mathbf{X}, \mathbf{P}, \mathbf{C}) = \hat{R}_{k,m}(\mathbf{X}, \mathbf{P}, \mathbf{C}) - \bar{R}_{k,m}^p(\mathbf{X}, \mathbf{P}, \mathbf{C}), \\
		R_{k,m}(\mathbf{X}, \mathbf{P}, \mathbf{C}) = \frac{1}{|\mathcal{K}_m|}  R_{k,m}^c(\mathbf{X}, \mathbf{P}, \mathbf{C}) + R_{k,m}^p(\mathbf{X}, \mathbf{P}, \mathbf{C}).	
	\end{cases}
\end{equation}
\subsection*{B. UAV positioning}
At the \((t+1)\)-th iteration, given the the power allocation result \(\mathbf{P}^{(t)} = \{p_{j,m}^{(t)}\}\) and association matrix \(\mathbf{C}^{(t)} = \{c_{k,m}^{(t)}\}\), for the $t$-th iteration, we first solve the UAV localization subproblem. We adopt the BCD and SCA technique to construct a concave surrogate function around the current local point \(\mathbf{X}^{(t)} = \{x_{m}^{(t)}\}\) for this iteration.
For the channel gain ${g}_k(\mathbf{x}_m)$, a non-concave form is:
$
\tilde{g}_k(\mathbf{x}_m; \mathbf{x}_m^t) \overset{\Delta}{=} \beta_k(\mathbf{x}_m^t) \|\mathbf{x}_m - \mathbf{u}_k\|^{-\alpha_k(\mathbf{x}_m^t)},
$
Since $\tilde{g}_k$ is not concave w.r.t. $\mathbf{x}_m$, a concave approximation is derived using the first-order Taylor expansion w.r.t. $\|\mathbf{x}_m - \mathbf{u}_k\|^2$ as follows:
\begin{equation}
	\begin{aligned}
		\tilde{g}_k(\mathbf{x}_m; \mathbf{x}_m^t) &\approx \phi_{k,m}^t \left( \|\mathbf{x}_m^t - \mathbf{u}_k\|^2
		- \|\mathbf{x}_m - \mathbf{u}_k\|^2  \right) \\
		& \quad  + \tilde{g}_k(\mathbf{x}_m^t; \mathbf{x}_m^t),
	\end{aligned}
\end{equation}
where:
$
\phi_{k,m}^t = \frac{\alpha_k(\mathbf{x}_m^t) \beta_k(\mathbf{x}_m^t)}{2 \|\mathbf{x}_m^t - \mathbf{u}_k\|^{2 + \alpha_k(\mathbf{x}_m^t)}}.
$
 
The concave approximations of the rate functions $\hat{R}_{k,m}$, $\bar{R}_{k,m}^c$, and $\bar{R}_{k,m}^p$ around $\mathbf{X}^t$ are derived via first-order Taylor expansion, replacing $g_k(\mathbf{x}_m)$ by $\tilde{g}_k(\mathbf{x}_m; \mathbf{x}_m^t)$. The resulting approximations with respect to $\hat{R}_{k,m}$, $\bar{R}_{k,m}^c$ and $\bar{R}_{k,m}^p$ are defined as follows:

\begin{equation}
	\footnotesize
	\begin{aligned}
		&\hat{R}_{k,m}(\mathbf{X}, \mathbf{P}^t, \mathbf{C}^t) \\
		&\approx \theta_{k,m} \sum_{m \in \mathcal{M}} \phi_{k,m}^t 
		\left( \frac{p_{c,m}^t}{\sigma^2} + \sum_{j \in \mathcal{K}_m} 
		\frac{p_{j,m}^t}{\sigma^2} \right) \\
		&\quad \cdot \left( \|\mathbf{x}_m^t - \mathbf{u}_k\|^2 
		- \|\mathbf{x}_m - \mathbf{u}_k\|^2 \right) 
		+ \hat{R}_{k,m}(\mathbf{X}^t, \mathbf{P}^t, \mathbf{C}^t) \\
		&\overset{\Delta}{=} \hat{R}_{k,m}^{\text{appr}}(\mathbf{X}; 
		\mathbf{X}^t, \mathbf{P}^t, \mathbf{C}^t),
	\end{aligned}
\end{equation}
\begin{equation}
	\footnotesize
	\begin{aligned}
		&\bar{R}_{k,m}^c(\mathbf{X}, \mathbf{P}^t, \mathbf{C}^t) \\
		&\approx \theta_{k^*,m}^c \left[ \phi_{k^*,m}^t \sum_{j \in \mathcal{K}_m} \frac{p_{j,m}^t}{\sigma^2} 
		\left( \|\mathbf{x}_m^t - \mathbf{u}_{k^*}\|^2 - \|\mathbf{x}_m - \mathbf{u}_{k^*}\|^2 \right) \right. \\
		&\quad + \sum_{m' \in \mathcal{M} \setminus \{m\}} \phi_{k^*,m'}^t 
		\left( \frac{p_{c,m'}^t}{\sigma^2} + \sum_{j \in \mathcal{K}_{m'}} \frac{p_{j,m'}^t}{\sigma^2} \right) \\
		&\quad \left. \cdot \left( \|\mathbf{x}_{m'}^t - \mathbf{u}_{k^*}\|^2 - \|\mathbf{x}_{m'} - \mathbf{u}_{k^*}\|^2 \right) \right] 
		+ \bar{R}_{k,m}^c(\mathbf{X}^t, \mathbf{P}^t, \mathbf{C}^t) \\
		&\overset{\Delta}{=} \bar{R}_{k,m}^{c,\text{appr}}(\mathbf{X}; \mathbf{X}^t, \mathbf{P}^t, \mathbf{C}^t),
	\end{aligned}
\end{equation}
\begin{equation}
	\footnotesize
	\begin{aligned}
		&\bar{R}_{k,m}^p(\mathbf{X}, \mathbf{P}^t, \mathbf{C}^t) \\
		&\approx \theta_{k,m}^p(\mathbf{X}^t,\mathbf{P}^t) \left[
		\phi_{k,m}^t \sum_{\substack{j \in \mathcal{K}_m,  j > k}} \frac{p_{j,m}^t}{\sigma^2} 
		\left( \|\mathbf{x}_m^t - \mathbf{u}_k\|^2 - \|\mathbf{x}_m - \mathbf{u}_k\|^2 \right) \right. \\
		&\quad + \sum_{m' \in \mathcal{M} \setminus \{m\}} \phi_{k,m'}^t 
		\left( \frac{p_{c,m'}^t}{\sigma^2} + \sum_{j \in \mathcal{K}_{m'}} \frac{p_{j,m'}^t}{\sigma^2} \right) \\
		&\quad \left. \cdot \left( \|\mathbf{x}_{m'}^t - \mathbf{u}_k\|^2 - \|\mathbf{x}_{m'} - \mathbf{u}_k\|^2 \right) \right] 
		+ \bar{R}_{k,m}^p(\mathbf{X}^t, \mathbf{P}^t, \mathbf{C}^t) \\
		&\overset{\Delta}{=} \bar{R}_{k,m}^{p,\text{appr}}(\mathbf{X}; \mathbf{X}^t, \mathbf{P}^t, \mathbf{C}^t),
	\end{aligned}
\end{equation}
respectively. The corresponding coefficients are:
\begin{equation}
	\footnotesize
	\begin{cases}
		\begin{aligned}
			\theta_{k,m} &= \frac{c_{k,m}^t\cdot \ln(2)^{-1}}{\left(1 + P_{k,\text{total}}(\mathbf{X}^t, \mathbf{P}^t)\right)}, \\
			\theta_{k^*,m}^c &= \frac{c_{k^*,m}^t\cdot \ln(2)^{-1}}{\left(1 + \sum_{j \in \mathcal{K}_m} \frac{p_{j,m}^t}{\sigma^2} g_{k^*}(\mathbf{x}_m^t) + I_{k^*,m}^{\text{other}}(\mathbf{X}^t, \mathbf{P}^t, m)\right)}, \\
			\theta_{k,m}^p &= \frac{c_{k,m}^t\cdot \ln(2)^{-1}}{\left(1 + \sum_{\substack{j \in \mathcal{K}_m, j > k}} \frac{p_{j,m}^t}{\sigma^2} g_k(\mathbf{x}_m) + I_{k,m}^{\text{other}}(\mathbf{X}^t, \mathbf{P}^t, m)\right)}.
		\end{aligned}
	\end{cases}
\end{equation}

Therefore, the concave function approximation of $R_{k,m}(\mathbf{X}, \mathbf{P}, \mathbf{C})$ can be expressed by:
\begin{equation}
	\small
	\begin{aligned}
		&{R}_k^{\text{dep}}(\mathbf{X}; \mathbf{X}^t, \mathbf{P}^t, \mathbf{C}^t) \\
		&= \sum_{m \in \mathcal{M}} \Bigg[ 
		\frac{1}{|\mathcal{K}_m|} \left( \hat{R}_{k,m}^{\text{appr}}(\mathbf{X}; \mathbf{X}^t, \mathbf{P}^t, \mathbf{C}^t) 
		- \bar{R}_{k,m}^{c,\text{appr}}(\mathbf{X}; \mathbf{X}^t, \mathbf{P}^t, \mathbf{C}^t) \right)\\ 
		&+ \hat{R}_{k,m}^{\text{appr}}(\mathbf{X}; \mathbf{X}^t, \mathbf{P}^t, \mathbf{C}^t) 
		- \bar{R}_{k,m}^{p,\text{appr}}(\mathbf{X}; \mathbf{X}^t, \mathbf{P}^t, \mathbf{C}^t) 
		\Bigg].		
	\end{aligned}
\end{equation}
Since $\rho (\Lambda^t, \mathbf{C}^t)$ is a constant, it can be omitted during the optimization of UAVs' positions, so that only position-related constraints are considered as follows:
\begin{align}\label{eq:optimal_X}
	\small
	\underset{\mathbf{X}}{\text{max}} \quad &\underset{k \in \mathcal{K}}{\text{min}}  {R}_k^{\text{dep}}(\mathbf{X}; \mathbf{X}^t, \mathbf{P}^t, \mathbf{C}^t)\\
	\text{s.t.} \quad &	\|\mathbf{x}_j - \mathbf{x}_j^{(t)}\| \leq \zeta^{(t)},\\
	&\eqref{eq:optimization_g}. \notag
\end{align}

The radius of the spherical region, denoted by $\zeta^{(t)}$, is gradually reduced during the iteration process to ensure convergence. A practical update rule is given by $\zeta^{(t+1)} = \eta \zeta^{(t)}$, where $\eta < 1$ is the step size. With this approximation, the optimization problem becomes concave that can be efficiently solved using standard convex tools such as Convex Optimization Modeling System (CVX). 
\subsection*{C. Power and Association}
Next, we solve the UAV power allocation and user association subproblem. Following the method for UAV positioning, we construct a concave surrogate function at the local point \((\mathbf{P}^{(t)}, \mathbf{C}^{(t)})\), and denote the optimal solution to problem \eqref{eq:optimal_X} as \(\mathbf{X}^{(t+1)}\). Given the updated UAV positions \(\mathbf{X}^{(t+1)} = \{\mathbf{x}_m^{(t+1)}\}\), the surrogate function \(R_{k,m,n}(\mathbf{X}^{(t+1)}, \mathbf{P}, \mathbf{C})\) is constructed in the \((t+1)\)-th iteration as:
\begin{small}
	\begin{align}
		&R_{k,m}(\mathbf{X}^{t+1}, \mathbf{P}, \mathbf{C}) 
		\approx R_{k,m}(\mathbf{X}^{t+1}, \mathbf{P}, \mathbf{C}^t)  \\
		&\hspace*{1em} + R_{k,m}(\mathbf{X}^{t+1}, \mathbf{P}^t, \mathbf{C}) 
		- R_{k,m}(\mathbf{X}^{t+1}, \mathbf{P}^t, \mathbf{C}^t). \notag
	\end{align}
\end{small}

Note that $R_{k,m}(\mathbf{X}^{t + 1}, \mathbf{P}, \mathbf{C}^t)=\hat{R}_{k,m}(\mathbf{X}^{t + 1}, \mathbf{P}, \mathbf{C}^t) - \tilde{R}_{k,m}(\mathbf{X}^{t + 1}, \mathbf{P}, \mathbf{C}^t)$ is the difference of two concave functions w.r.t. $\mathbf{P}$. Since any concave function is globally upper-bounded by its first-order Taylor expansion at any point, we have:
\begin{equation}
	\footnotesize
	\begin{aligned}
		&\bar{R}_{k,m}^c(\mathbf{X}^{t + 1}, \mathbf{P}, \mathbf{C}^t) \leq \theta_{k^*,m}^{'c} \Bigg[ 
		\sum_{j \in \mathcal{K}_m} \frac{g_{k^*}(\mathbf{x}_{m'}^{t + 1})}{\sigma^2} (p_{j,m} - p_{j,m}^t) \\  
		&\hspace{1em} + \sum_{m' \in \mathcal{M} \setminus \{m\}} \Bigg( 
		\frac{g_{k^*}(\mathbf{x}_{m'}^{t + 1})}{\sigma^2} (p_{c,m'} - p_{c,m'}^t) \\
		&\hspace{1em} + \sum_{j \in \mathcal{K}_{m'}} \frac{g_{k^*}(\mathbf{x}_{m'}^{t + 1})}{\sigma^2} (p_{j,m'} - p_{j,m'}^t) 
		\Bigg) \Bigg] + \bar{R}_{k,m}^c(\mathbf{X}^{t + 1}, \mathbf{P}^t, \mathbf{C}^t) \\
		&\overset{\Delta}{=} \bar{R}^{c,\text{upper}}_{k,m}(\mathbf{P}; \mathbf{X}^{t + 1}, \mathbf{P}^t, \mathbf{C}^t),
	\end{aligned}
\end{equation}
\begin{equation}
	\footnotesize
	\begin{aligned}
		&\bar{R}_{k,m}^p(\mathbf{X}^{t + 1}, \mathbf{P}, \mathbf{C}^t) \leq \theta_{k,m}^{'p} \Bigg[
		\sum_{\substack{j \in \mathcal{K}_m, j > k}} \frac{g_k(\mathbf{x}_m^{(t + 1)})}{\sigma^2} (p_{j,m} - p_{j,m}^t) \\
		&\hspace{1em} + \sum_{m' \in \mathcal{M} \setminus \{m\}} \Bigg(
		\frac{g_k(\mathbf{x}_{m'}^{(t + 1)})}{\sigma^2} (p_{c,m'} - p_{c,m'}^t) \\
		&\hspace{1em} + \sum_{j \in \mathcal{K}_{m'}} \frac{g_k(\mathbf{x}_{m'}^{(t + 1)})}{\sigma^2} (p_{j,m'} - p_{j,m'}^t) 
		\Bigg) \Bigg] + \bar{R}_{k,m}^p(\mathbf{X}^{t + 1}, \mathbf{P}^t, \mathbf{C}^t) \\
		&\overset{\Delta}{=} \bar{R}^{p,\text{upper}}_{k,m}(\mathbf{P}; \mathbf{X}^{t + 1}, \mathbf{P}^t, \mathbf{C}^t),
	\end{aligned}
\end{equation}
with
\begin{equation}
	\footnotesize
	\begin{cases}
		\begin{aligned}
			\theta_{k,m}^{'c} = \frac{c_{k^*,m}^{t} \cdot \ln(2)^{-1}}{
				1 + \sum_{j \in \mathcal{K}_m} \frac{p_{j,m}^{t+1}}{\sigma^2} g_{k^*}(\mathbf{x}_m^t) + 
				I_{k^*,m}^{\text{other}}(\mathbf{X}^t,\mathbf{P}^t,m)},\\
			\theta_{k,m}^{'p} = \frac{c_{k,m}^{t} \cdot \ln(2)^{-1}}{
				1 + \sum_{\substack{j \in \mathcal{K}_m, j>k}} \frac{p_{j,m}^{t+1}}{\sigma^2} g_k(\mathbf{x}_m) + 
				I_{k,m}^{\text{other}}(\mathbf{X}^{t+1},\mathbf{P}^t,m)}.
		\end{aligned}
	\end{cases}	
\end{equation}

Hence, $R_{k,m}(\mathbf{X}^{t+1}, \mathbf{P}, \mathbf{C})$ is further approximated as:
\begin{equation}
	\footnotesize
	\begin{aligned}
		&R_{k,m}(\mathbf{X}^{t+1}, \mathbf{P}, \mathbf{C}^t) \\
		&\approx \frac{1}{|\mathcal{K}_m|}\left(\hat{R}_{k,m}(\mathbf{X}^{t+1}, \mathbf{P}, \mathbf{C}^t) 
		- \bar{R}^{c,\text{upper}}_{k,m}(\mathbf{P}; \mathbf{X}^{t+1}, \mathbf{P}, \mathbf{C}^t)\right) \\
		&\hspace{1em} + \hat{R}_{k,m}(\mathbf{X}^{t+1}, \mathbf{P}, \mathbf{C}^t) 
		- \bar{R}^{p,\text{upper}}_{k,m}(\mathbf{P}; \mathbf{X}^{t+1}, \mathbf{P}, \mathbf{C}^t) \\
		&\hspace{1em} + R_{k,m}(\mathbf{X}^{t+1}, \mathbf{P}^t, \mathbf{C}) 
		- R_{k,m}(\mathbf{X}^{l+1}, \mathbf{P}^t, \mathbf{C}^t) \\
		&\overset{\Delta}{=} R^{\text{appr}}_{k,m}(\mathbf{P}, \mathbf{C}; \mathbf{X}^{t+1}, \mathbf{P}^t, \mathbf{C}^t).
	\end{aligned}
\end{equation}

In addition, since $c_{k,m,n}^2 \geq 2c_{k,m,n}^t c_{k,m,n}-(c_{k,m,n}^t)^2$ holds for a given local point $c_{k,m,n}^t$, we have the following lower bound on $\rho(\boldsymbol{\Lambda}^t,\mathbf{C})$:
\begin{equation}
	\footnotesize
	\begin{aligned}
		&\rho(\mathbf{\Lambda}^t, \mathbf{C}) = -\sum_{k \in \mathcal{K}} \sum_{m \in \mathcal{M}} \lambda_{k,m}^t c_{k,m} (1 - c_{k,m}) \\
		&\qquad \geq \sum_{k \in \mathcal{K}} \sum_{m \in \mathcal{M}} \lambda_{k,m}^t \big( c_{k,m}^t(2 c_{k,m} - c_{k,m}^t) - c_{k,m} \big) \triangleq \rho^{\text{lb}}(\mathbf{C}; \mathbf{\Lambda}^t, \mathbf{C}^t).
	\end{aligned}
\end{equation}

Then, power and association optimization problems can be relaxed as the following convex problem:
\begin{align}\label{eq:optimal_PC}
	\footnotesize
	\underset{\mathbf{P}, \mathbf{C}}{\text{max}} \quad &\min_{k \in \mathcal{K}} \left( \sum_{m \in \mathcal{M}} R_{k,m}^{\text{pow,asso}}(\mathbf{P}, \mathbf{C}; \mathbf{X}^{t+1}, \mathbf{P}^t, \mathbf{C}^t) \right)\\
	& \hspace*{1em} + \rho^{\text{lb}}(\mathbf{C}; \mathbf{\Lambda}^t, \mathbf{C}^t)\notag\\
	\text{s.t.} \quad &\eqref{eq:optimization_c},\eqref{eq:optimization_d}, \eqref{eq:optimization_e}, \eqref{eq:optimization_f}, \eqref{eq:penalty_a}.\notag
\end{align}

Subsequently, the strong or weak association parameter for the (t+1)-th iteration is discretized by rounding as follows:
\begin{equation}\label{eq:roundC}
	\footnotesize
	\mathbf{C}^{t+1} = 
	\begin{cases} 
		0, & \mathbf{C}^{t+1} \leq 0.05, \\
		1, & \mathbf{C}^{t+1} \geq 0.95.
	\end{cases}
\end{equation}

Then the penalty multiplier is updated by using the following formula:
\begin{align}
	\footnotesize
	\lambda_{k,m}^{L+1} &= \lambda_{k,m}^t + \gamma^t {c}_{k,m}^{t+1} (1 - {c}_{k,m}^{t+1}),
\end{align}
\begin{align}
	\footnotesize
	\gamma^t &= -\frac{\mu^t}{\sum_{k \in \mathcal{K}} \sum_{m \in \mathcal{M}} \left({c}_{k,m}^{t+1} (1 - {c}_{k,m}^{t+1}) \right)^2}, 
\end{align}
with $\mu^{(t+1)} \gets 2 \mu^{(t)}$ at each iteration.

Based on the above analysis, the details of the algorithm is summarized in Algorithm 1. The algorithm terminates if the iteration number exceed $t_{max}$ and obtain the solution.
\begin{algorithm}[t]
	{\footnotesize
		\caption{Joint positioning, power allocation and  user association optimization}
		\textbf{Input:} The coordinates of the vertices of the buildings, $\{u_k, \forall k \in \mathcal{K} \}$, $P_{max}$, $\sigma^2$, $\alpha_1$, $\beta_1$, $\alpha_2$, $\beta_2$, $\eta$, $\zeta^{(0)}$, $\lambda_{k,m}^{(0)}$, $t_{max}$.\\
		\textbf{Output:} $\mathbf{X}^*$, $\mathbf{P}^*$, $\mathbf{C}^*$, $R^*$.
		\begin{algorithmic}[1]
			\State Calculate the LoS regions $\{\mathcal{D}_{k,q}, k \in \mathcal{K}, q \in \mathcal{Q} \}$.
			\State Set $t = 0$, and initialize $\mathbf{X}^{(0)}$, $\mathbf{P}^{(0)}$, and $\mathbf{C}^{(0)}$.
			\Repeat
			\State Update $\mathbf{X}^{(t+1)}$ by solving problem \eqref{eq:optimal_X}.
			\State Update $\mathbf{P}^{(t+1)}$ and $\mathbf{C}^{(t+1)}$ by solving problem \eqref{eq:optimal_PC}. And $\mathbf{C}^{(t+1)}$ rounds off according to \eqref{eq:roundC}.
			\State Update $\zeta^{(t+1)} \gets \eta ~\zeta^{(t)}$, $\mu^{(t+1)} \gets 2 \mu^{(t)}$.
			\State Update $t \gets t + 1$.
			\Until{The iteration number $t > t_{max}$.}
			\State $\mathbf{X}^* \gets \mathbf{X}^{(t)}$, $\mathbf{P}^* \gets \mathbf{P}^{(t)}$, $\mathbf{C}^* \gets \mathbf{C}^{(t)}$, $R^* \gets \min\limits_{k \in \mathcal{K}} R_k$.
			\State \textbf{return} $\mathbf{X}^*$, $\mathbf{P}^*$, $\mathbf{C}^*$, $R^*$
		\end{algorithmic}
	}
\end{algorithm}
\section{Performance evaluation}


In this section, we present simulation results to evaluate the performance of the proposed algorithm. The simulation is carried out  in an 800×800 $m^2$ area, the left half of the map features a high density of tall buildings, whereas the right half exhibits a similarly high density but with lower structures. The building geometries are predefined, and key simulation parameters are summarized in table below. 
To prevent collisions, the minimum UAV flight altitude is set to 101$m$. Initially, each user associates with the UAV offering the highest channel gain, UAVs are deployed at an altitude of 300m that positioned directly above one of $\mathcal{K}_m$ selected users.

\begin{table}[h]
	\centering
	\caption{Simulation Parameters}
	\begin{tabular}{cc}
		\specialrule{1pt}{1pt}{1pt}
		\textbf{Parameter} & \textbf{Value} \\
		\midrule
		Path loss exponent & $\alpha_{\text{LoS}} = 2,\ \alpha_{\text{NLoS}} = 3.3$ \\
		\multirow{2}{*}{Channel gain per meter} & $\beta_{\text{LoS}} = -46.43~\mathrm{dB}$ \\
		& $\beta_{\text{NLoS}} = -56.43~\mathrm{dB}$ \\
		Noise power density (5 MHz) & $N_0 = -107~\mathrm{dBm/Hz}$ \\
		Initial penalty factor & $\lambda_{k,m}^{[0]} = 0.05$ \\
		Initial radius of spherical region & $\zeta^{[0]} = 50~\mathrm{m}$ \\
		Radius reduction ratio & $\eta = 0.9$ \\
		Maximum iteration number & $t_{\max} = 15$ \\
		\specialrule{1pt}{1pt}{1pt}
	\end{tabular}
\end{table}

We first illustrate the proposed solution for UAV positioning, power allocation and user association. As shown in Fig. 3, the UAV trajectories are optimized to mitigate the path loss induced by obstacles. Compared to the initial state, the minimal transmission rate increases significantly from 0.3885 to 3.9364 bit/s/Hz, demonstrating the effectiveness of the proposed approach.
Fig. 4 presents a performance comparison between RSMA and NOMA transmission schemes across different users. In the final positioning, the UAVs serve an average of 3, 4, and 6 users, respectively. The results clearly show that RSMA consistently outperforms NOMA in terms of achievable user data rates.

Compared with the three baselines: fixed position, fixed power, and no geographic information, the proposed method achieve the best performance.
As the number of users increases, UAVs need to fly higher to maintain LoS connections with the users. In this case, the increased distance not only lead to severe propagation loss, but also causes interferences to non-targeted users.
It is worth noting that the fixed-position scheme performs worse than the fixed-power scheme. This is because UAVs positioned at higher altitudes cause stronger interferences to other users. without position optimization, they cannot use obstacles to block interference to non-targeted users.
The scheme that does not utilize environment information performs the worst, since any blockage can lead to significantly reduced transmission rates.
\begin{figure}[t]
	\centering
	\includegraphics[width=0.45\textwidth]{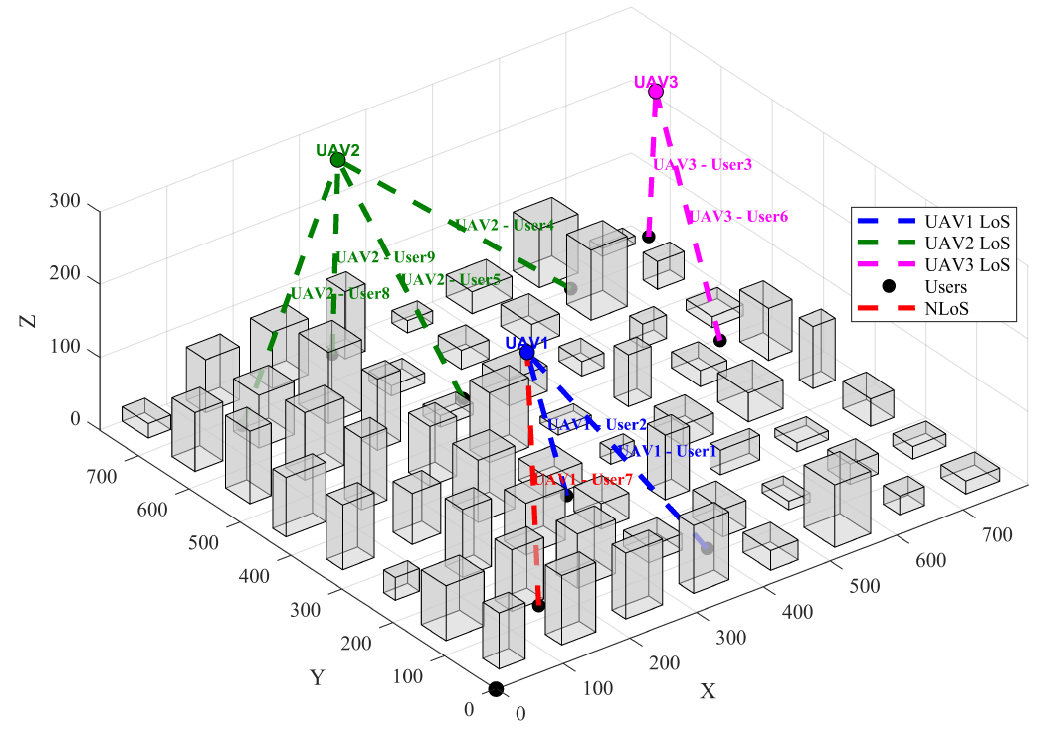}
	\hfill
	\includegraphics[width=0.45\textwidth]{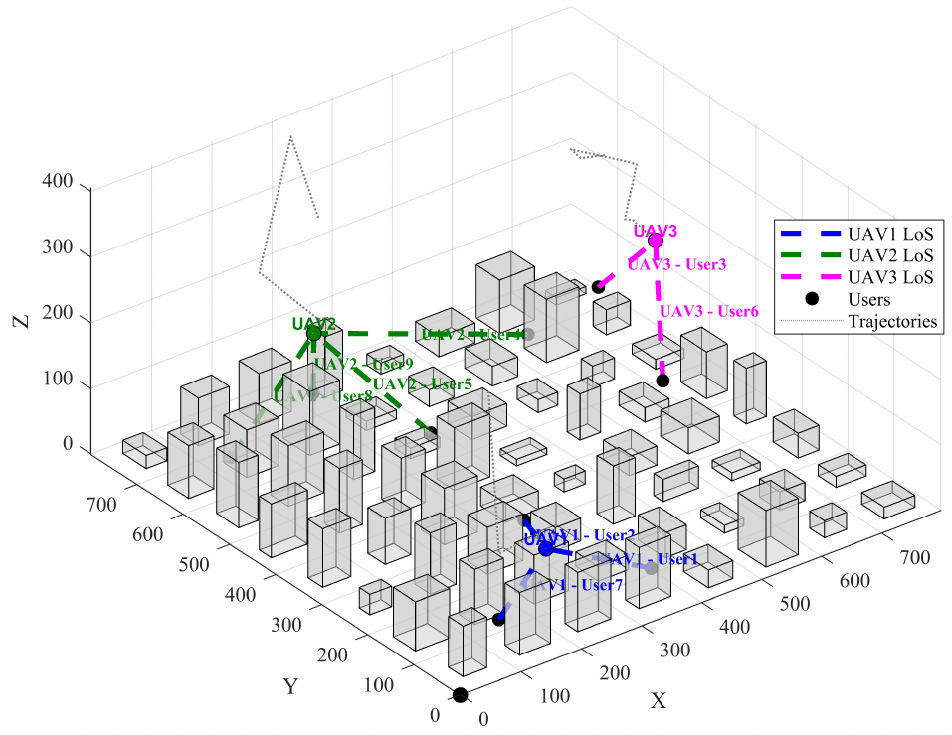}
	\caption{Demonstration the initial and final situations of UAVs after optimization.}
	\label{fig:uav_comparison}
\end{figure}

\begin{figure}[t]
	\centering
	\includegraphics[width=0.38\textwidth]{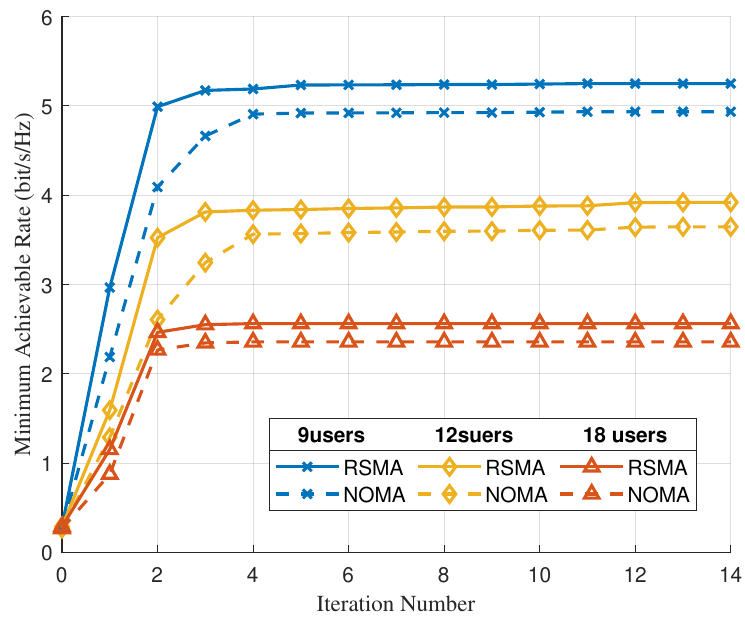}
	\caption{Convergence of minimum rate with three UAVs under RSMA and NOMA for varying user numbers.}
	\label{fig:picture3}
\end{figure}
\begin{figure}[t]
	\centering
	\includegraphics[width=0.39\textwidth]{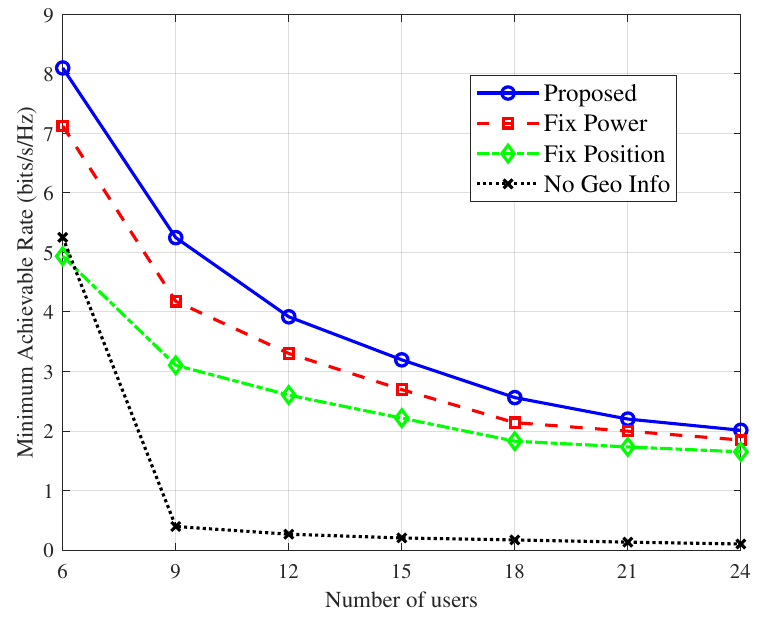}
	\caption{Minimum achievable rates for different schemes versus the number of users.}
	\label{fig:picture4}
\end{figure}
\section{Conclusion}
In this paper, we proposed a RSMA-based communication framework for multiple UAVs-assisted system to serve ground users while leveraging environmental obstacles to mitigate inferences. To address the complexity of the non-convex joint optimization problem involving UAV positioning, power allocation and user association, we introduced a penalty function-based relaxation strategy and developed an efficient iterative algorithm. The proposed approach can effectively maximize the minimum user rate by ensuring LoS links for served users and suppressing inferences through obstacle-free position optimization. Simulation results demonstrated that our method not only achieves reliable LoS connectivity and superior user rates compared to baselines optimizing only position or power, but also significantly outperforms the conventional NOMA scheme under the same environmental conditions.

\vspace{12pt}
\color{red}

\end{document}